\documentclass[rmp,showpacs,twocolumn]{revtex4}

\usepackage{graphicx}

\begin{document}

\title{Switching of sub-$\mu$m sized, antiferromagnetically coupled CoFeB/Ru/CoFeB trilayers}

\author{N. Wiese}
\email[Electronic Mail: ]{mail@nilswiese.de}

\affiliation{Siemens AG, Corporate Technology, Paul-Gossen-Str. 100,
91052 Erlangen, Germany} \affiliation{University of Bielefeld, Nano
Device Group,
 Universit\"atsstr. 25, 33615 Bielefeld, Germany}

\author{T. Dimopoulos}
\affiliation{Siemens AG, Corporate Technology, Paul-Gossen-Str. 100,
91052 Erlangen, Germany}

\author{M. R\"uhrig}
\affiliation{Siemens AG, Corporate Technology, Paul-Gossen-Str. 100,
91052 Erlangen, Germany}

\author{J. Wecker}
\affiliation{Siemens AG, Corporate Technology, Paul-Gossen-Str. 100,
91052 Erlangen, Germany}

\author{G. Reiss}
\affiliation{University of Bielefeld, Nano Device Group,
Universit\"atsstr. 25, 33615 Bielefeld, Germany}

\date{\today}

\begin{abstract}
This work reports on the magnetic reversal of sub-$\mu$m sized
elements consisting of an CoFeB/Ru/CoFeB artificial ferrimagnet
(AFi). The elements were patterned into ellipses having a width of
approximately 250 to 270nm and a varying aspect ratio between 1.3
and 8. The coercivity was found to decrease with an increasing
imbalance of the magnetic moment of the two antiferromagnetically
coupled layers and is therefore strongly affected by an increase of
effective anisotropy due to the antiferromagnetic coupling of the
two layers. With respect to a single layer of amorphous CoFeB,
patterned in comparable elements, the AFi has an increased
coercivity. Switching asteroids comparable to single layers were
only observed for samples with a high net moment.
\end{abstract}


\maketitle


\section{Introduction}
Magnetic tunnel junctions (MTJ) have gained considerable interest in
recent years due to their high potential in various applications,
e.g. as reads heads \cite{Ho01}, angle \cite{Berg99} or strain
sensors \cite{Loehndorf02a} and as programmable resistance in data
storage (MRAM) \cite{Gallagher97} or even magnetic logic devices
\cite{Richter02a}.

\begin{figure}
    \includegraphics[width=7cm]{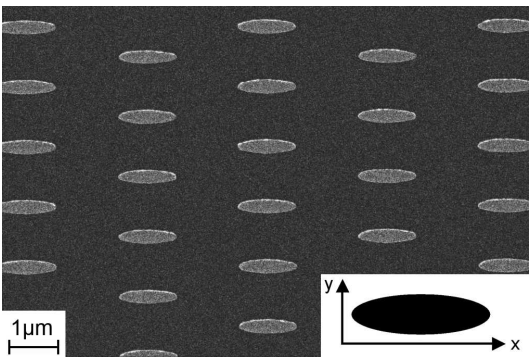}
    \caption{SEM images of one of the arrays of sample D, showing ellipses with
the size of $0.27\mu \mbox{m} \times 1.18 \mu \mbox{m}$.}
    \label{fig:SEM}
\end{figure}

The underlaying concept is a spin valve consisting of a hard
magnetic reference electrode separated from the soft magnetic sense
or storage layer by a tunnel barrier like Al$_2$O$_3$. The reference
layer usually is an artificial ferrimagnet (AFi) exchange biased by
a natural antiferromagnet, in which the AFi consists of two
ferromagnetic layers coupled antiparallel via a thin non-magnetic
spacer. For the soft electrode, mostly single layers of
polycrystalline material, e.g NiFe and CoFe, have been used
\cite{Koch98}. Recently, soft electrodes of polycrystalline AFis,
based on ferromagnetic materials like CoFe and NiFe, have been
investigated. They show a further reduction of the stray field due
to the reduced net moment, smaller switching field distribution
\cite{Sousa02} and an easier establishment of a single domain
structure in patterned elements with small aspect ratio
\cite{Tezuka03a}.

Additionally, the concept of an AFi allows one to further adjust the
magnetic properties of the soft layer. Compared to the coercivity of
a continuous single layer, $H_{\mathbf c}^{\mathbf{SL}}$, the
coercivity of the AFi, $H_{\mathbf c}^{\mathbf{AFi}}$, is enhanced
by a factor $Q$:
\begin{eqnarray}
H_{\mathbf c}^{\mathbf {AFi}} = Q \cdot H_{\mathbf c}^{\mathbf {SL}} \quad \mbox{with} \quad Q = \frac{M_1 t_1 + M_2 t_2}{M_1 t_1 - M_2 t_2} \label{Qvalue}
\end{eqnarray}
where $M_1, M_2$ and $t_1, t_2$ are the saturation magnetization and
the thickness of the two composite ferromagnetic
layers.\cite{Berg96} As shown before for unpatterned AFi films
consisting of two amorphous, ferromagnetic CoFeB layers, separated
by a thin Ru spacer, the coercivity can be tailored in a wide range
and is approximately by a factor of nine smaller than in systems of
polycrystalline CoFe/Ru/CoFe.\cite{Wiese05} Furthermore the coupling
of the amorphous CoFeB-AFi shows an oscillating behavior in
dependence of the thickness of the nonmagnetic Ru spacer and
achieves a coupling strength of the order of $-0.1
\mbox{mJ}/\mbox{m}^2$ at the second antiferromagnetic maximum, which
is about a factor of ten smaller than in polycrystalline
CoFe/Ru/CoFe trilayers.\cite{Wiese04}


It was the purpose of this study to investigate the switching
behavior of the amorphous CoFeB-AFi at sub-$\mu$m sizes, where
additional shape anisotropy and the magnetostatic edge coupling have
to be taken into account. These contributions lead to an {\em
effective anisotropy} which is for patterned elements different from
the anisotropy of continuous films (eqn. \ref{Qvalue}).

As the amorphous alloy we chose Co$_{60}$Fe$_{20}$B$_{20}$ due to a
high tunnel magnetoresistance (TMR) \cite{Wang04} and an enhanced
temperature stability of the TMR \cite{Dimopoulos04}. A thin Ru
spacer was used to mediate the coupling between the two
ferromagnetic layers of the AFi.

\begin{table}
\caption{\label{tab:table} Investigated samples and parameters
extracted from the AGM measurements of the unpatterned layer
systems.}
\begin{ruledtabular}
\begin{tabular}{ccccc}
 sample & system & $Q_{\mbox{\scriptsize meas}}$ & $H_{\mbox{\scriptsize sat}}$ & J \\
        &        &                    & $\lbrack \frac{\mbox{\scriptsize kA}}{\mbox{\scriptsize m}} \rbrack$ & $\lbrack \frac{\mbox{\scriptsize mJ}}{\mbox{\scriptsize m}^2} \rbrack$ \\ \hline
 A & CoFeB 3.5 / Ru 0.95 / CoFeB 3 & 7.7 & 29.8 & -0.06 \\
 B & CoFeB 4.0 / Ru 0.95 / CoFeB 3 & 5.1 & 26.3 & -0.06 \\
 C & CoFeB 4.5 / Ru 0.95 / CoFeB 3 & 3.7 & 23.9 & -0.06 \\
 D & CoFeB 4 & & & \\ \hline
\end{tabular}
\end{ruledtabular}
\end{table}


\section{Experimental}
Samples have been deposited by magnetron sputtering on thermally
oxidized SiO$_2$ wafers at a base pressure of $5 \cdot 10^{-8}$
mbar. A magnetic field of approximately 4 kA/m was applied during
deposition in order to induce the easy axis in the magnetic layers.
The AFi was grown on a 1.2 nm thick Al layer, oxidized in an
Ar/O${_2}$ plasma for 0.8 min without breaking the vacuum, to have
similar growth conditions as in a MTJ. In order to investigate the
switching behavior in dependence on the Q-value, three different
samples have been prepared where the thickness of the ferromagnetic
layer in contact with the Al$_2$O$_3$ layer, and thus the net
magnetic moment, has been varied (see table \ref{tab:table}, samples
A to C). Additional a single CoFeB layers with a thickness of 4 nm
(sample D) has been deposited for comparison. All samples were
capped with a Ta layer to protect the multilayers from oxidation.

The samples have been patterned by a single step e-beam lithography
and Ar-ion etching process. For lithography a positive e-beam resist
was used, leading to patterns with a small edge roughness and high
reproducibility across the whole array. On each sample different
arrays of ellipses with a nominal width, $w$, of 250nm and varying
length, $\ell$, have been defined, leading to different aspect
ratios, $u = \ell/w$, between 1.3 and 8. The lateral distances have
been chosen three times the dimension of the elements, and therefore
a dipolar coupling between the individual ellipses of an array can
be neglected.\cite{Abraham_mmm04} Each of the arrays had the
dimension of $25 \times 25 \mu \mbox{m}^2$. After the arrays have
been coated by a Ta layer of appropriate thickness (ranging from 8
to 15nm), the capping was removed in a lift-off process in a bath of
solvent under application of ultrasonic agitation. During etching
with a 80 $\mu A/cm^2$ Ar ion current the samples were tilted by
approximately 30 degrees and rotated to obtain a uniform etch
profile over the sample. The etching depth was monitored by a
secondary ion mass spectrometer (SIMS) attached to the etching
facility.

The sizes and the uniformity of all patterns have been characterized
by scanning electron microscopy (SEM) after the patterning process.
The SEM image in figure \ref{fig:SEM} shows one of the arrays of
sample D with ellipses in the size of $0.27 \mu \mbox{m} \times 1.18
\mu \mbox{m}$, confirming the high uniformity of the patterns. The
width for samples with small aspect ratios ($u = \ell/w < 5$) varied
between 250 and 270nm. Due to a tendency of over-exposure, ellipses
with larger aspect ratios show a slightly larger widths of
approximately 300nm.

After patterning all samples have been field annealed for 20min. at
150$^\circ \mbox{C}$ and $475\mbox{kA/m}$ applied parallel to the
long axis of the ellipses using a vacuum annealing furnance.
Magnetization loops of all arrays have been taken by a commercial
magneto-optical Kerr effect magnetometer (MOKE) with a typical spot
diameter of 4$\mu$m.\cite{NanoMOKE}

\begin{figure}
    \includegraphics[width=8.0cm]{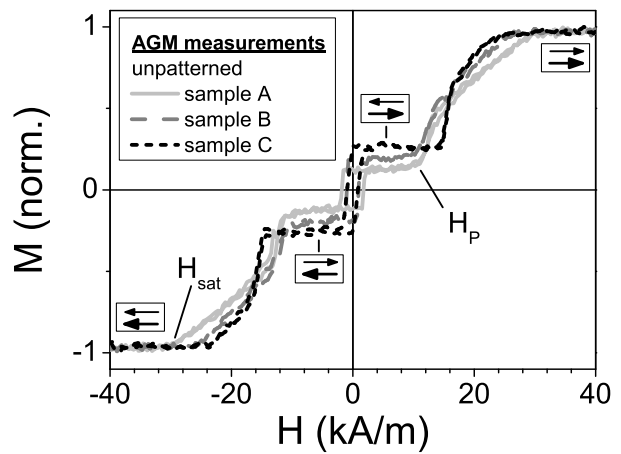}
    \caption{(color online). Magnetization loops of unpatterned samples taken by AGM, showing a
well established coupling for all AFi samples. The saturation field,
$H_{\mbox{\scriptsize sat}}$, and the plateau field,
$H_{\mbox{\scriptsize p}}$, are indicated for sample A.}
    \label{fig:agmloops}
\end{figure}

\section{Results and Discussion}
A room temperature magnetization curve, $M(H)$, of the
antiferromagnetically (AF) coupled systems is shown in figure
\ref{fig:agmloops}. From the $M(H)$ loops, obtained by alternating
gradient field magnetometery (AGM), one can extract the saturation
field, $H_{\mbox{\scriptsize sat}}$, the total,
$m_{\mbox{\scriptsize tot}}= m_1 + m_2$, and the net magnetic
moment, $m_{\mbox{\scriptsize net}}= m_1 - m_2$, allowing to
calculate the measured Q-value, $Q_{\mbox{\scriptsize meas}} =
m_{\mbox{\scriptsize tot}}/m_{\mbox{\scriptsize net}}$, the
individual magnetization of the layers, $m_{1,2}$, and the coupling
energy \cite{Berg96}
\begin{equation}
J=- \mu_0 H_{\mbox{\scriptsize sat}} \frac{m_1 m_2}{m_1+m_2}
\label{coupling}
\end{equation}
The coupling is $-0.06 \mbox{mJ}/\mbox{m}^2$ for all investigated
AFi samples. The values are in accordance to the oscillating
coupling in dependence on the Ru spacer thickness around the second
antiferromagnetic maximum as presented elsewhere.\cite{Wiese04} The
measured Q-values vary between 3.7 and 7.7 for the investigated AFi
samples, and are significantly smaller than the $Q$ value,
$Q_{\mbox{\scriptsize nom}}$, calculated from the nominal thickness
of the ferromagnetic layers. This discrepancy is supposed to result
from a thicker magnetically dead layer of the upper CoFeB layer in
comparison to the bottom layer. Since the samples are well protected
from oxidation, as confirmed by Auger depth profiling, this would
indicate a stronger intermixing of the upper CoFeB interfaces. All
data extracted from the AGM measurements are given in table
\ref{tab:table}.

\begin{figure}
    \includegraphics[width=7cm]{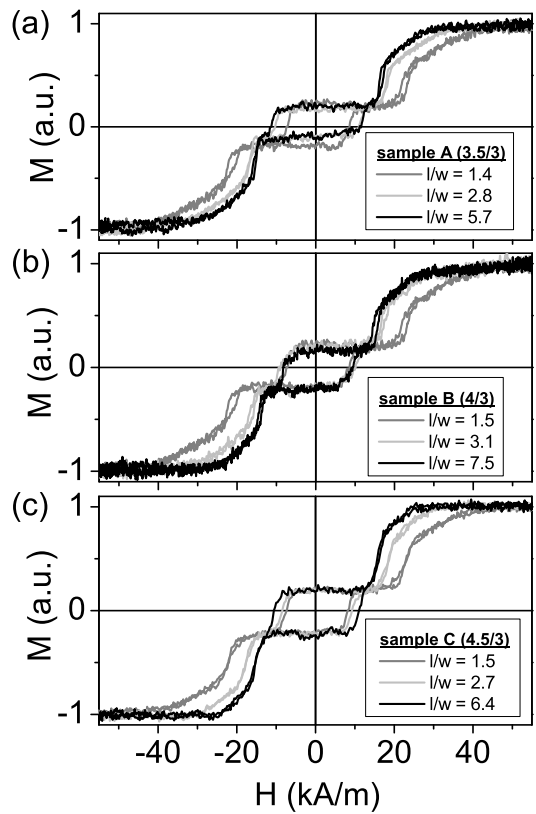}
    \caption{(color online). Major magnetization loops measured by MOKE of some of the patterned
arrays, showing a well established coupling and
$H_{\mbox{\scriptsize p}} > H_{\mbox{\scriptsize c}}$ for all
samples.}
    \label{fig:MajorloopsPatterned}
\end{figure}

Figure \ref{fig:MajorloopsPatterned} shows the magnetization loops
of patterned arrays of all AFi samples as obtained by MOKE
measurements. The strong AF coupling is maintained after the
patterning and annealing steps. Additionally, the saturation field
is increasing with decreasing aspect ratio (i.e. length or size) of
the ellipses, due to an increase in stray field coupling between the
layers within the AFi system. For large aspect ratios
$H_{\mbox{\scriptsize sat}}$ achieves the values of the unpatterned
samples.

The saturation field can be expressed by two contributions, one
originating from the antiferromagnetic interlayer coupling, the
other resulting from the stray field coupling. Whereas the first
should depend on ${ - \frac{J}{\mu_0} \frac{m_1 + m_2}{m_1 m_2}}$,
as derived from eqn. \ref{coupling}, the latter should depend on
$\mu_0 M_s \frac{t_{\mbox{\scriptsize tot}}}{w} n_x$,\cite{Sun01}
where $t_{\mbox{\scriptsize tot}}$ is the total thickness of the
AFi. The second contribution only depends on the demagnetization
factor $n_x$, since the $y$-components in case of an AF coupled
system are compensated for external fields larger than the plateau
field, $H_{\mbox{\scriptsize p}}$. Since the width of the elements
was hold constant for the investigated samples, the dependence of
$H_{\mbox{\scriptsize sat}}$ on the sample geometry is only given by
the demagnetization factor\cite{Hubert98}
\begin{eqnarray}
n_x=\frac{u}{2} \int_0^{\infty} \frac{ds}{(u^2+s) \cdot
\sqrt{(u^2+s)(1+s)s}}
\end{eqnarray}

\begin{figure}
    \includegraphics[width=8cm]{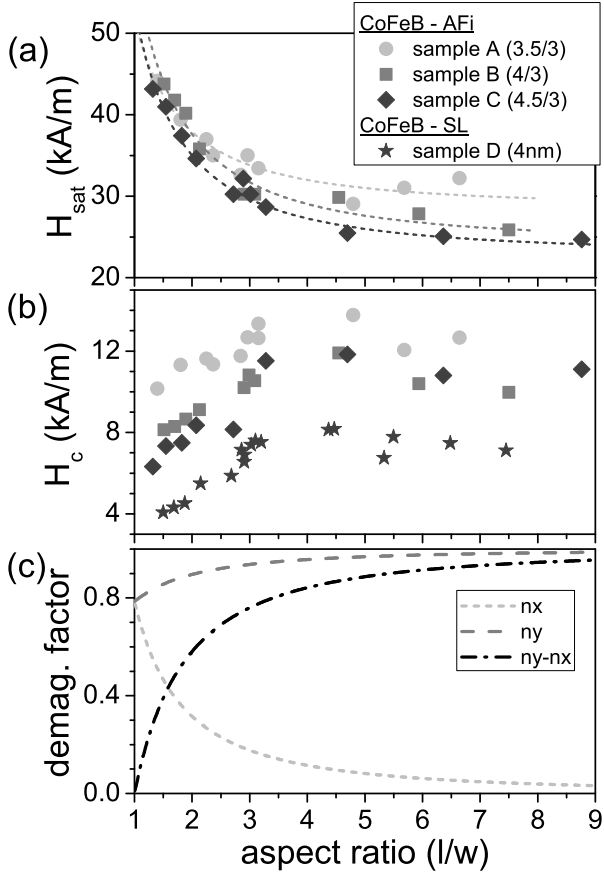}
    \caption{(color online). (a) Saturation field, H$_{\mbox{\scriptsize sat}}$, and (b)
coercivity, H$_{\mbox{\scriptsize c}}$, of patterned samples vs. the
aspect ratio $u=\ell/w$. The width of approximately 250 to 270nm was
kept constant for all investigated samples. The dashed lines in (a)
represent a fit with a function proportional to the demagnetization
factor $n_{\mbox{\scriptsize x}}$. (c) Numerical calculated
demagnetization factors for ellipses against the aspect ratio.}
    \label{fig:HsatHc}
\end{figure}

This model is fitted to the measured data as shown in figure
\ref{fig:HsatHc}(a). The fitting shows a high accordance between the
experimental data and the model, thus verifying the dependence of
the saturation field on $n_{\mbox{\scriptsize x}}$ for small aspect
ratios. From the fitting parameters one can further extract the
saturation field for an infinite elongated ellipse to 28.6 kA/m for
sample A, 24.2 kA/m for sample B and 22.8 kA/m for sample C,
respectively. This values are in good agreement to the measured
saturation fields at the unpatterned AFi samples given by the
interlayer coupling (see table \ref{tab:table}).

For external fields smaller than $H_{\mbox{\scriptsize p}}$, the AF
coupling remains stable. As can be seen from figure
\ref{fig:MajorloopsPatterned}, the AFi should reverse its
magnetization like a single layer sample with a reduced net moment
and enhanced effective anisotropy. Therefore it is possible to
measure minor magnetization loops in a small field window ($\pm 10$
to $\pm15$ kA/m, depending on $H_{\mbox{\scriptsize p}}$ of the
sample) and extract the coercivity, $H_{\mbox{\scriptsize c}}$, of
the arrays (see figure \ref{fig:HsatHc}(b)).

For small aspect ratios ($u<4$), $H_{\mbox{\scriptsize c}}$
increases with $u$ by approximately 3.5 kA/m and remains constant
for large aspect ratios (see figure \ref{fig:HsatHc}(b)). The slight
decrease in coercivity for the largest aspect ratios is most likely
attributed to the slightly larger width of the ellipses. This
behavior of the coercivity vs. aspect ratio holds also for the
single layer sample. The increase of $H_{\mbox{\scriptsize c}}$ for
small aspect ratios is caused by the increase in shape anisotropy,
which for an ellipse is given by\cite{Sun01}
\begin{eqnarray}
H_{\mbox{\scriptsize k}} = 4 \pi M_{\mbox{\scriptsize
s}}\frac{t_{\mbox{\scriptsize net}}}{w}(n_y - n_x) \label{Sun}
\end{eqnarray}
where the demagnetization factor ($n_y - n_x$) depends on the aspect
ratio. Therefore the experimental results of figure \ref{fig:HsatHc}
(b) are in qualitative accordance with the ($n_y - n_x$) dependence
on aspect ratio (see figure \ref{fig:HsatHc}(c)). Deviations from
the calculated dependence for large aspect ratios are likely to
result from micromagnetic differences: Small elements are stabilized
by a non vanishing stray field, arising from the magnetic poles of
the elements. These stray fields stabilize the overall magnetization
of the elements, so that the reversal can be more accurately
approximated by a single domain mechanism. For large aspect ratios
the magnetic poles are larger separated, thus minimizing the stray
field coupling and as a result a nucleation driven magnetization
reversal, most likely initiated by edge domains, is more favourable,
leading to a almost constant coercivity.\cite{Meyners03}

Unlike elliptic elements with a single ferromagnetic layer, the
coercivity of the patterned AFi samples does not scale proportional
to the net magnetic moment. With higher net moment, the coercivity
is {\it decreasing}, but remaining always larger than for elements
of a 4nm thick single layer. The reason is a superposition of the
effective anisotropy due to the AF coupling ($\sim
m_{\mbox{\scriptsize tot}}/m_{\mbox{\scriptsize net}}$), as
expressed by eqn. \ref{Qvalue} for the unpatterned films, and the
dependence of coercivity on the net magnetic moment as described by
eqn. \ref{Sun} (${\sim m_{\mbox{\scriptsize net}} (n_y - n_x)/w}$).
Overall the influence of the AF coupling is dominating for the
investigated CoFeB AFis and the coercivity is increased by
approximately a factor of 1.4 when decreasing the net thickness of
the AFi, and therefore the net moment, from 1.5 to 0.5nm.

\begin{figure}
    \includegraphics[width=7cm]{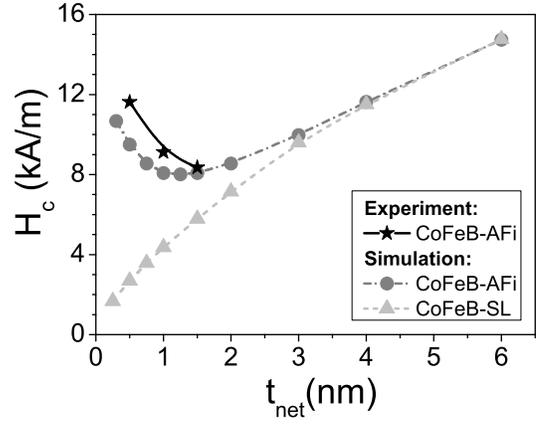}
    \caption{Experimental and calculated dependence of coercivity,
$H_{\mbox{\scriptsize c}}$, on the net thickness,
$t_{\mbox{\scriptsize net}}$, of the AFi and a single layer,
respectively. The experimental data points are for elliptical shaped
elements with the dimension of $250\mbox{nm} \times 520\mbox{nm}$,
the simulated data are evaluated for the same sample geometry. Lines
are guides to the eye.}
    \label{fig:simulation}
\end{figure}

The dependence of coercivity on the net thickness,
$t_{\mbox{\scriptsize net}} = t_2 - t_1$, of the AFi and for single
layer samples with layer thicknesses of $t=t_{\mbox{\scriptsize
net}}$ have been simulated within a micromagnetic model using
Landau-Lifshitz-Gilbert equations (see figure
\ref{fig:simulation}).\cite{Scheinfein} For the simulation of the
AFis the thickness of the second FM layer, $t_2$, was varied between
3.5 and 6nm, whereas the thickness of the first FM layer was kept
fixed to $t_1=3\mbox{nm}$, and the thickness of the nonmagnetic
spacer was chosen to 1nm. The saturation moment of the FM layers
were assumed to $M_{\mbox{\scriptsize s}}=860 \mbox{emu/cm}^3$, the
uniaxial anisotropy to $K_{\mbox{\scriptsize u}}=2.4 \cdot 10^{3}
\mbox{erg/cm}^3$, the exchange stiffness constant to $A=1.05 \cdot
10^{-6} \mbox{erg/cm}$ and for the AF coupling constant to $J=-0.004
\cdot 10^{-6} \mbox{erg/cm}$. The coercivity of the simulated
elliptic AFi elements of $250\mbox{nm} \times 520\mbox{nm}$
decreases with net thickness for $t_{\mbox{\scriptsize
net}}\leq1.5$nm, and increases for larger $t_{\mbox{\scriptsize
net}}$, asymptotically reaching the values of the single layer
coercivity. The experimental data of the AFi arrays with the same
geometry show a similiar behavior, therefore verifying the above
described model for the samples under study.

\begin{figure}
    \includegraphics[width=8.6cm]{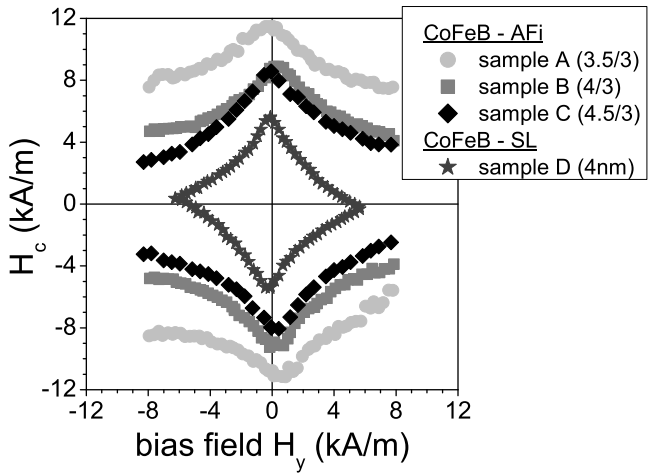}
    \caption{(color online). Bias field dependence of the coercivity for all investigated film
systems with an aspect ratio $u \approx 2.2$, illustrating also the
gain in effective anisotropy by introduction of the
antiferromagnetic coupling and the strong influence of the increased
Q-value on the anisotropy of the system.}
    \label{fig:Asteroids}
\end{figure}

Figure \ref{fig:Asteroids} shows the bias field dependence of the
coercivity for samples patterned with an aspect ratio of $u \approx
2.2$. Due to the increase of effective anisotropy with the Q-value
(and therefore basically with the inverse net magnetic moment), the
asteroid gets stretched along the hard axis field. If used in a
conventional writing scheme as a soft magnetic electrode for
applications like MRAM or magnetic logic, the broad asteroid shape
of the AFi storage layer cells limits the choice of Q-value due to
field limitations, reducing the proposed advantages of an AFi
structure as reported by others.\cite{Sousa02,Tezuka03a}

\section{Conclusion}
The AFi based on an amorphous CoFeB alloy shows a strong increase in
the effective ansiotropy due to the AF coupling, mediated by a thin
Ru interlayer and by the stray field coupling, respectively. This is
reflected by a higher coercivity and an increase of the asteroid
width compared to a single CoFeB layer. The dependence of the
switching field on the net magnetic moment can not be explicitely
explained within the model that considers the AFi as a rigid
ferromagnetic layer with a reduced moment. One has further to take
into account the increase of effective anisotropy, which basically
scales with the inverse net moment for the investigated combinations
of layer thicknesses. This last factor appears to be dominating in
the system under study and leads to the decrease of coercivity with
net moment. Finally, it has been found, that the saturation field of
the patterned AFis is decreasing with larger aspect ratio and is
asymptotically reaching the saturation field of the unpatterned
films. The origin for this behavior for small aspect ratios can be
found in the additional contribution of the stray field coupling of
the two ferromagnetic layers within the AFi. Therefore the
antiparallel alignment of the system is additionally favoured by the
stray field coupling, depending basically on the demagnetization
factor $n_{\mbox{\scriptsize x}}$ vs. aspect ratio.

\section*{Acknowledgments}
The authors wish to thank J. Bangert and G. Gieres for fruitful
discussions, H. Mai and K. Rott for experimental support. Financial
support of the German Ministry for Education and Research is
gratefully acknowledged (Grant no. 13N8208).



\begin{thebibliography}{00}

\bibitem{Ho01} M.K. Ho, C.H. Tsang, R.E. Fontana, S.S.P. Parkin, K.J. Carey, T. Pan, S. MacDonald, P.C. Arnett, and J.O. Moore, IEEE Trans. Magn. {\bf 37}, 1691
(2001)

\bibitem{Berg99} H.A.M. van den Berg, J. Altmann, L. B\"ar, G. Gieres, R. Kinder, G. Rupp, M. Vieth, and J. Wecker, IEEE Trans. Magn. {\bf 35}, 2892
(1999)

\bibitem{Loehndorf02a} M. L\"ohndorf, T. Duenas, M. Tewes, E. Quandt, M. R\"uhrig, and J. Wecker , Appl. Phys. Lett. {\bf 81}, 313
(2002)

\bibitem{Gallagher97} W.J. Gallagher, S.S.P. Parkin, Y. Lu, X.P. Bian, A. Marley, K.P. Roche, R.A. Altman, S.A. Rishton, C. Jahnes, T.M. Shaw, and G. Xiao, J. Appl. Phys. {\bf 81}, 3741
(1997)

\bibitem{Richter02a} R. Richter, L. B\"ar, J. Wecker, and G. Reiss, Appl. Phys. Lett. {\bf 80}, 1291
(2002)

\bibitem{Koch98} R.H. Koch, J.G. Deak, D.W. Abraham, P.L. Trouilloud, R.A. Altman, Y. Lu, W.J. Gallagher, R.E. Scheuerlein, K.P. Roche, and S.S.P. Parkin, Phys. Rev. Lett. {\bf 81}, 4512
(1998)

\bibitem{Sousa02} R.C. Sousa, Z. Zhang, and P.P. Freitas, J. Appl. Phys. {\bf 91}, 7700
(2002)

\bibitem{Tezuka03a} N. Tezuka, N. Koike, K. Inomata, and S. Sugimoto, J. Appl. Phys. {\bf 93}, 7441
(2003)

\bibitem{Berg96} H.A.M van den Berg, W. Clemens, G. Gieres, G. Rupp, and M. Vieth , IEEE Trans. Magn. {\bf 32}, 4624
(1996)

\bibitem{Wiese05} N. Wiese, T. Dimopoulos, M. R\"uhrig, J. Wecker, H. Br\"uckl, and G. Reiss, J. Magn. Magn. Mater. {\bf 290-291}, 1427
(2005)

\bibitem{Wiese04} N. Wiese, T. Dimopoulos, M. R\"uhrig, J. Wecker, H. Br\"uckl, and G. Reiss, Appl. Phys. Lett. {\bf 85}, 2020
(2004)

\bibitem{Wang04} D. Wang, C. Nordman, J. Daughton, Z. Qian, and J. Fink, IEEE Trans Magn {\bf 40}, 2269
(2004)

\bibitem{Dimopoulos04} T. Dimopoulos, N. Wiese, G. Gieres, J. Wecker and M.D. Sacher, J. Appl. Phys. {\bf 96},
6382 (2004)

\bibitem{Abraham_mmm04} D.W. Abraham, and Y. Lu, J. Appl. Phys. {\bf 98}, 023902 (2005)

\bibitem{NanoMOKE} for detailed specification of the NanoMOKE2\texttrademark system look at the website of Durham Magneto Optics Ltd., http://durhammagnetooptics.com

\bibitem{Sun01} J.Z. Sun, J.C. Slonczewski, P.L. Trouilloud, D. Abraham, I. Bacchus, W.J. Gallagher, J. Hummel, Y. Lu, G. Wright, S.S.P. Parkin, and R.H. Koch, Appl. Phys. Lett. {\bf 78}, 4004 (2001)

\bibitem{Hubert98} A. Hubert, R. Sch\"afer, {\it Magnetic Domains, The Analysis of Magnetic Microstructures}, Springer
(1998)

\bibitem{Meyners03} D. Meyners, H. Br\"uckl, and G. Reiss, J. Appl. Phys. {\bf 93}, 2676 (2003)

\bibitem{Scheinfein} The commercial available micromagnetic program (LLG Micromagnetics Simulator\texttrademark)
developed by M.R. Scheinfein was used, see
http://llgmicro.home.mindspring.com

\end{thebibliography}
\end{document}